\documentclass[a4paper]{article}

\begin{document}
\title{\bf Curvature force and dark energy}
\author{
Alexander B. Balakin\footnote{Electronic address: Alexander.Balakin@ksu.ru}\\
{\small Department of General Relativity and Gravitation}  \\
{\small Kazan State University, 420008 Kazan, Russia} \\
Diego Pav\'{o}n\footnote{Electronic address:
Diego.Pavon@uab.es}\\
{\small Departamento de F\'{\i}sica, Universidad Aut\'{o}noma de Barcelona}\\
{\small 08193 Bellaterra (Barcelona), Spain}\\
Dominik J. Schwarz\footnote{Electronic address: dominik.schwarz@cern.ch}\\
{\small Theory Division, CERN, 1211 Geneva 23, Switzerland}\\
and\\
Winfried Zimdahl\footnote{Electronic address:
zimdahl@thp.uni-koeln.de}\\
{\small Fachbereich Physik, Universit\"at Konstanz and} \\
{\small Institut f\"ur Theoretische Physik, Universt\"at zu K\"oln,
50937 K\"oln, Germany}}

\date{February 7, 2003; May 14, 2003 (revised)}
\maketitle

\begin{abstract}
A curvature self-interaction of the cosmic gas is shown to mimic a
cosmological constant or other forms of dark energy, such as a
rolling tachyon condensate or a Chaplygin gas. Any given Hubble
rate and deceleration parameter can be traced back to the action
of an effective curvature force on the gas particles. This force
self-consistently reacts back on the cosmological dynamics. The
links between an imperfect fluid description, a kinetic
description with effective antifriction forces, and curvature
forces, which represent a non-minimal coupling of gravity to
matter, are established.
\end{abstract}


\section{Introduction}
\label{introduction}

An adequate description of the present Universe seems to require a
cosmic substratum, which is characterized by a negative pressure
\cite{Riess}. In particular observations of supernovae at high
redshift strongly suggest that the Universe is accelerating its
expansion \cite{RT}. A possible explanation is the existence of a
dominant component of dark energy, besides cold dark matter
(pressureless). There exist a number of dark energy candidates,
the best known being a cosmological constant and different
quintessence scenarios (\cite{RaPeeb,caldwell,amendola,ZP}). Most
of the latter rely on the dynamics of a minimally coupled scalar
field. But also non-minimal ``extended quintessence'' models have
been studied, which are characterized by an explicit coupling of
the scalar field to the Ricci scalar \cite{Matar}. This additional
coupling results in a richer dynamical structure of the theory,
which has been used to search for scaling and tracker field
solutions \cite{Uzan}. A different type of non-minimal coupling is
obtained from higher-order theories of gravity, which was shown to
give rise to the concept of  ``curvature quintessence''
\cite{Capo}. Geometric terms in fourth-order gravity are
interpreted as effective quantities within general relativity such
as ``curvature pressure'' and ``curvature density''. Under certain
conditions the curvature pressure may be sufficiently negative to
generate a phase of accelerated expansion.  This kind of
modification of the gravitational action was previously used in
connection with different problems, as for instance to avoid the
initial singularity of homogeneous, isotropic universes
\cite{Kerner}.

A negative pressure may also be the consequence of
self-interactions in gas models of the Universe
\cite{ZiBa01,ZSBP,ZBEntr,S}. In particular, an ``antifrictional''
force, self-consistently exerted on the particles of the cosmic
substratum, was shown to provide an alternative explanation for an
accelerated expansion of the universe \cite{ZSBP,S}. This approach
relies on the fact that the cosmological principle is compatible
with the existence of a certain class of (hypothetical)
microscopic one-particle forces, which manifest themselves as
``source'' terms in the macroscopic perfect-fluid balance
equations. These sources can be mapped on an effective negative
pressure of the cosmic medium. The energy-momentum tensor of the
latter thus acquires an imperfect fluid structure. An advantage of
this approach is the possibility to unify dark energy and dark
matter, since just a single dark component has to be introduced to
describe cosmological observations. However, a compelling
microphysical explanation for antifrictional forces is still
missing, as is the case for all other models.

The energy-momentum tensor of a non-minimally coupled scalar field
has an imperfect fluid structure as well \cite{Madsen}, which
reduces to that of a perfect fluid in the limit of minimal
coupling. This indicates that there might be a relation between
imperfect fluid degrees of freedom and non-minimal coupling. Here
we exploit the general idea of describing a non-minimal coupling
within an imperfect fluid picture.
We point out that effective antifrictional forces can be regarded
as a specific non-minimal coupling of the cosmic gas to the Ricci
scalar. Generally, a force that explicitly depends on curvature
quantities describes a coupling of matter to the space-time
curvature, which goes beyond Einstein's theory. However, mapping
the non-minimal interaction on an imperfect fluid degree of
freedom admits a self-consistent treatment on the basis of general
relativity. This may be seen as a gas dynamical counterpart to the
non-minimal couplings of scalar fields or those of higher-order
gravity theories. We emphasize that the present coupling
represents a new type of curvature self-interaction of the cosmic
medium, which cannot be reduced to just a mapping of the described
scalar field approaches to a fluid description. The starting
points are quite different. The non-minimal scalar field
approaches start with a given interaction term and then look for
suitable solutions for the cosmological dynamics. It is not clear
from the outset which coupling could provide a ``successful''
solution.

Here, we use an inverse strategy. We {\it design} a (non-minimal)
fluid interaction such that it results in the desired cosmic
evolution. Designing the coupling to obtain a specific dynamics
has already been used for interacting two-component models
\cite{ZP}. We apply this idea to the case of a one-component
fluid, which is self-consistently coupled to the Ricci scalar. As
a characteristic feature of this approach, Hubble rate and
deceleration parameter explicitly enter the microscopic dynamics,
which gives rise to a self-consistent coupling of the latter to
the gravitational field equations. We demonstrate this for a
power-law behaviour of the scale factor, implying a specifically
rolling tachyon field, for the $\Lambda$CDM model and for a
(generalized)  Chaplygin gas. All these cases may be understood as
the result of specific curvature self-interactions in an otherwise
pressureless gas.

A problem common to all unified models (dark energy and dark
matter being the same component) is that a tiny, non-vanishing
speed of sound can spoil the scenario by inducing acoustic
oscillations of primordial, adiabatic fluctuations at late times
\cite{Sandvik,Bean}. The fact that neither oscillations nor
exponential instabilities are observed, at scales of galaxy
clusters and below, puts severe constraints on the isentropic
speed of sound in such models, i.e. $c^2_{\rm s} < 10^{-5}$
\cite{Sandvik}. In fact, this seems to exclude any perfect
fluid model, which does not mimic a $\Lambda$CDM model. As a
consequence, the Chaplygin gas, say, cannot be considered a
realistic model of the cosmic substratum. One should be aware,
however, that the mentioned limits are derived under the
assumption of an equation of state $P=P(\rho)$, where $P$ is the
total pressure and $\rho$ is the energy density. Since a general
(dissipative) fluid has to be described by an equation of state of
the type $P=P(\rho,s)$, where $s$ is the (specific) entropy, it
remains open whether or not these constraints apply in this more
general case as well.  The point is that dissipative processes in
imperfect fluids give rise to entropy perturbations and we have
$c_{\rm s}^2 \neq \dot{P}/\dot{\rho}$. This implies that a simple
relation between the perturbations of pressure and energy density,
which was used to obtain the constraints in \cite{Sandvik,Bean},
does not necessarily exist.

The paper is organized as follows. In section \ref{Field equations
and viscous pressure} we relate an effective, non-equilibrium type
pressure to the Ricci scalar of a homogeneous and isotropic,
spatially flat Universe. A gas dynamical motivation for this
pressure as the result of a non-minimal curvature self-interaction
is given in section \ref{Kinetic theory and curvature
self-interaction}. In section \ref{Curvature force and accelerated
expansion} the mentioned examples, a power-law behaviour, including
a special case of a rolling tachyon, the $\Lambda$CDM model, and
the Chaplygin gas are considered. A brief summary is given in
section \ref{Conclusions}. Units are fixed by $c = k_{B} = h =1$.

\section{Field equations and viscous pressure}
\label{Field equations and viscous pressure}

The field equations for a homogeneous, isotropic, and spatially
flat Universe filled by an imperfect fluid are
\begin{equation}
3 H^2 = 8\pi G \rho\ , \qquad
\dot{H} = - 4\pi G\left(\rho + P\right)\ , \qquad
P = p + \Pi  \ .
\label{1}
\end{equation}
Here, $\rho$ is the energy density seen by a comoving observer.
The fluid  four-velocity $u^i$ is normalized by $u^i u_i=-1$.   
The Hubble rate is given by $H = \dot{a}/a$, where $a$
is the scale factor of the Robertson-Walker metric and a dot
denotes a derivative with respect to cosmic time $t$. The pressure
$P$ of the cosmic medium is assumed to be the sum of a kinetic
part $p > 0$ [see Eq.~(\ref{12}) below] and an additional contribution
$\Pi$. The derivation of the latter quantity from the type of
self-interactions mentioned in the introduction is the main
objective of the paper. We shall show that such a pressure appears
as the result of an effective one-particle force $F^{i}$ of structure
$mF^i = B\left(- E p^i + m^2 u^i\right)$ [see Eq.~(\ref{17})
below], where $m$ is the mass of the gas particles, $p^i$ is
their four momentum and $E \equiv -u_ip^i$ is the particle energy
for a comoving (with the macroscopic four velocity) observer. The
pressure $\Pi$ will directly be related to the force function $B$
[see Eqs.~(\ref{19}) and (\ref{20}) below].

{}From the definition, $q \equiv  - \ddot{a}/(aH ^{2})$, of the
deceleration parameter, one has
\begin{equation}
q = - 1 - \frac{\dot{H}}{H ^{2}}\ .
\label{3}
\end{equation}
For the special case of a constant  $q > -1$, we find
\begin{equation}
H = \frac{1}{\left(1 + q\right)t}\ , \quad a(t) \propto t^{\frac{1}{1+q}}\ ,
\label{4}
\end{equation}
and
\begin{equation}
H = H_0 \ , \qquad a(t) = a_0 e^{H_0 t}\ ,
\label{5}
\end{equation}
for $q = -1$. Often, an accelerated expansion of the Universe is
traced back to a suitably designed scalar field potential. There
exist other approaches, which imply a non-minimal coupling of a
scalar field to the Ricci scalar \cite{Matar,Capo}.

Here, we obtain an accelerating expansion of the Universe within a
fluid picture, due to a sufficiently large negative effective
non-equilibrium pressure $\Pi$.  A conventional bulk viscous
pressure of linear irreversible thermodynamics is inappropriate
for this purpose, since it corresponds to a fluid configuration
which is close to a fiducial equilibrium reference state such that
the total pressure is positive. Non-standard self-interactions of
the cosmic medium, however, have been considered as a potential
mechanism to generate an accelerated expansion
\cite{ZiBa01,ZSBP,ZBEntr}.  In this paper we demonstrate how
these interactions can be obtained as the result of a non-minimal
coupling of the underlying gas dynamics to the space-time
curvature. For this purpose it is convenient to solve
Eqs.~(\ref{1}) with (\ref{3}) for $P/\rho$, which yields
\begin{equation}
\frac{P}{\rho} =  {1\over 3}\left(- 1 + 2 q\right)
\ .
\label{6}
\end{equation}
In general, $q$ is time-dependent.
The ratio $P/\rho$ may be related to the Ricci scalar, which in a
homogeneous, isotropic, and spatially flat Universe is given by
\begin{equation}
R = 6 \left(\frac{\ddot{a}}{a} + \frac{\dot{a}^{2}}{a ^{2}}\right)
= 6 \left(1 - q \right)H ^{2}\ .
\label{8}
\end{equation}
Therefore we may also write
\begin{equation}
\frac{P}{\rho } = -{1\over 3}
\left[\frac{R}{3H ^{2}} - 1\right]\ .
\label{9}
\end{equation}
The simplest way of obtaining the latter relation is to combine
the trace of Einstein's equation, $-R = 8\pi GT = - 8\pi G (\rho -
3P)$, with Friedmann's equation, $8\pi G\rho = 3H^2$. For the
Einstein--de Sitter Universe one has $a \propto t ^{2/3}$ and $R =
3 H ^{2}$, equivalent to $q={1 \over 2}$, i.e. $P = 0$. In the
following we shall look for a mechanism that produces deviations
from $R = 3 H ^{2}$, leading to a negative pressure. {}From now on
we focus on non-relativistic matter and set $p = 0 $, thus
$P = \Pi$. This is motivated by observations of the large scale structure,
which suggest that a non-relativistic equation of state ($p \ll \rho$) 
is required at the onset of structure formation and thereafter. 

\section{Kinetic theory and curvature self-interaction}
\label{Kinetic theory and curvature self-interaction}

Equation (\ref{9}) shows that to have a non-vanishing dissipative
pressure $\Pi$, a departure from $R = 3H^2$ is necessary. This
comes about because $R = 3 H ^{2}$ characterizes a perfect fluid
Universe with the equation of state for dust. Before focusing on
such departures, we recall that from a gas dynamical point of view
a perfect fluid consists of particles with mass $m$, which move on
geodesics according to
\begin{equation}
m\frac{\mbox{d} x ^{i}}{\mbox{d} \tau} = p ^{i}\ , \qquad
\frac{\mbox{D} p ^{i}}{\mbox{d} \tau} = 0\ .
\label{10}
\end{equation}
The parameter $\tau$ denotes the proper time. This corresponds to a Boltzmann
equation for the one-particle distribution function $f = f\left(x,p\right)$
(see, e.g. \cite{Ehl,Stew,IS,Groot}),
\begin{equation}
p^{i}f,_{i} - \Gamma^{i}_{kl}p^{k}p^{l}
\frac{\partial f}{\partial
p^{i}}
 = C\left[f\right]  \mbox{ , }
\label{11}
\end{equation}
where $C[f]$ is Boltzmann's collision integral. The latter describes elastic
binary collisions between the particles. The second moment of the distribution
function provides us with the energy-momentum tensor, which, in a spatially
homogeneous and isotropic Universe, has necessarily a perfect fluid
structure, i.e.
\begin{equation}
T^{ik} = \int \mbox{d}P p^{i}p^{k}f \left(x,p\right)
= \rho u^{i}u^{k} + p h^{ik}\ ,
\label{12}
\end{equation}
where $h^{ik} = g^{ik} + u^i u^k$. The continuity equation
$\dot{\rho} + 3H\left(\rho + p\right) =0$ follows, with a pressure
in the range $0 < p \leq \rho /3$. The special case of a dust
universe is approached for $p \ll \rho$, which is obtained for $T
\ll m$, where $T$ is the fluid equilibrium temperature
(\cite{Ehl,Stew,IS,Groot}). In particular, the  kinetic pressure
is always non-negative.

Our strategy now is the following. Under the assumption that a
gaseous fluid description makes sense, we attribute the
accelerated expansion of the Universe to the existence of a
non-vanishing dynamical pressure $P$ in Eq. (\ref{9}). We ask for
a suitable modification of the above perfect fluid description,
which might give rise to a negative pressure of a substantial
amount. A natural option for such a modification consists of additional
interparticle interactions, not taken into account by Boltzmann's
collision integral (e.g. inelastic interactions or many-particle
effects). The currently unknown properties of dark energy (and
dark matter as well) are then mapped onto non-standard
interactions between the microscopic constituents of the fluid.
That is, we shall look for those interactions that are able to
reproduce the observed cosmological dynamics. This strategy
resembles the more familiar scalar field approach according to
which one tries to ``explain'' the dynamics of the Universe by
designing a potential term in order to reproduce the given
dynamics.

Following previous work \cite{ZSBP}, we introduce additional
interactions, which cannot be reduced to elastic, binary
collisions. There are specific interactions, which may be mapped
onto a quantity $F ^{i}$ such that the Boltzmann equation
(\ref{11}) is generalized to
\begin{equation}
p^{i}f,_{i} - \Gamma^{i}_{kl}p^{k}p^{l}
\frac{\partial f}{\partial
p^{i}} + m F ^{i}\frac{\partial{f}}{\partial{p ^{i}}}
 = C\left[f\right]\ .
\label{15}
\end{equation}
The left-hand side of this equation can be regarded as
\[
\frac{\mbox{d}f \left(x,p \right)}{\mbox{d}\tau}
\equiv  \frac{\partial{f}}{\partial{x ^{i}}}
\frac{\mbox{d}x ^{i}}{\mbox{d}\tau}
+ \frac{\partial{f}}{\partial{p ^{i}}}
\frac{\mbox{d}p ^{i}}{\mbox{d}\tau}\ ,
\]
with
\begin{equation}
m\frac{\mbox{d} x ^{i}}{\mbox{d} \tau} = p ^{i}\ , \qquad
\frac{\mbox{D} p ^{i}}{\mbox{d} \tau} = F^i\ ,
\label{16}
\end{equation}
the equations of motion for gas particles moving  under the
influence of a force field $F ^{i}=F ^{i}\left(x,p \right)$. As a
consequence, the particle motion is no longer geodesic. However,
describing interactions in terms of a four-force raises the
question of the extent to which such a procedure is consistent with the
assumption of a spatially homogeneous and isotropic Universe.

To answer this question it is convenient to split the microscopic
particle momentum according to $p^i = Eu^i + \lambda e^i$, where
$u^i e_i = 0$ and $e^i e_i =1$. Here, $E \equiv - p^i u_i$ is the
particle energy as measured by an observer, comoving with the
macroscopic four-velocity $u ^{i}$. {}From $p^i p_i = - m^2$ we have
$E^2 = m^2 + \lambda^2$. In general, the individual particles do
not move with the mean velocity $u ^{i}$. Apparently, homogeneous
and isotropic models require a geodesic mean motion, but not
necessarily a geodesic motion of the individual particles. To
clarify the situation it is useful to introduce the {\it particle}
velocity $u_{\left(p\right)}^i$, defined by $p^i \equiv
mu_{(p)}^i$, which is not necessarily geodesic, and to contrast it
with the velocity $u^i$ of the geodesic mean motion. The particle
velocity is also normalized by $u_{\left(p\right)}^i
u_{\left(p\right)i} = -1$.

In order to get an idea about the admissible forces, it seems suggestive to
assume $F^i$ to be proportional to the difference $u^i - u_{\left(p\right)}^i$,
i.e. to start with an ansatz $F^i \propto u^i - u_{\left(p\right)}^i$.
On the other hand, the relation $F^i p_i = 0$ has to be satisfied. But the
latter condition, together with the ansatz $F^i \propto u^i -
u_{\left(p\right)}^i$, leads to $E=m$, the case that characterizes the mean
motion with $u_{\left(p\right)}^i = u^i$, which is force-free. It follows
that a non-vanishing force cannot simply be proportional to the difference
between the macroscopic and the particle velocities. A more general ansatz is
\[
\frac{F^i}{m} = B u^i - Cu^i_{\left(p\right)} \ ,
\]
where the quantities $B$ and $C$ are not constants but should depend on the
particle and fluid quantities in such a way that $B=C$ only for
$u_{\left(p\right)}^i = u^i$ in order to guarantee that the mean motion
remains force-free. With this ansatz we obtain
\[
F^ip_i = 0 \quad \Rightarrow \quad C = \frac{E}{m} B \ ,
\]
which indeed provides us with $C = B$ for $E=m$, equivalent to
$u_{\left(p\right)}^i = u^i$. For the force we find under such conditions
\begin{equation}
mF^i = B\left(- E p^i + m^2 u^i\right)
= - B u^k \left(g^i_k p^m p_m - p^i p_k\right)\ .
\label{17}
\end{equation}
The expression in the parenthesis on the right-hand side of the second
equation coincides with the projector orthogonal to the particle momentum.
In the special case $p^i =m u^i$, we have $E=m$ and the force consistently
vanishes. A force of the type (\ref{17}) makes the individual particles
move on non-geodesic trajectories, while the macroscopic mean motion remains
geodesic. This force, which was used in \cite{ZSBP}, is compatible with the
cosmological principle. Now we have to investigate whether, and under which
circumstances, a deviation from the geodesic motion of the microscopic
constituents due to a force (\ref{17}) may result in an effective negative
pressure of the cosmic medium. In the following we shall restrict ourselves
to the case where $B$ does {\it not} depend on $E$.

An interaction term in the Boltzmann equation gives rise to ``source'' terms
in the balances of the moments. In particular, from the balance for the
second moment of $f$ we obtain
\begin{equation}
\dot{\rho }+ 3H \left(\rho + p \right)
= -3B\left(\rho + p \right)\ .
\label{18}
\end{equation}
As before, $\rho$ and $p$ are defined by
\[
\rho = u_i u_k\int \mbox{d}P p^{i}p^{k}f \left(x,p\right)\quad {\rm and}\quad
p= \frac{1}{3}h_{ik}\int \mbox{d}P p^{i}p^{k}f \left(x,p\right)\ ,
\]
respectively. With the definition
\begin{equation}
\Pi H \equiv   B\left(\rho + p \right)
\ ,
\label{19}
\end{equation}
the energy balance (\ref{18}) becomes
\begin{equation}
\dot{\rho } + 3 H\left(\rho + p + \Pi \right)
= 0 \ .
\label{20}
\end{equation}
This proves that, macroscopically, the action of the force
manifests itself as a  dissipative pressure. The reinterpretation
of the right-hand side of Eq.~(\ref{18}) in terms of an effective
pressure is crucial for our approach. It maps the source in the
energy balance, which is a consequence of the additional
interaction, onto an imperfect fluid degree of freedom of a
conserved energy-momentum tensor $T^{ik}_{\rm eff} = \rho u^i u^k
+ \left(p + \Pi\right)h^{ik}$. We emphasize that the quantity
$\Pi$ does {\it not} coincide with the dissipative pressure of
conventional, linear, irreversible fluid dynamics. The latter has
its origin in Boltzmann's collision integral and may provide only
small corrections  in   $p$. Here, it is the force (\ref{17}),
which, via the identification (\ref{19}), generates an effective
pressure of an entirely different kind. There is {\em no}
restriction of the type $|\Pi|<p$, which is characteristic of
conventional fluid dynamics.

In the following, we are interested in  $\Pi \leq 0$, equivalent
to $B\leq 0$. We assume the cosmic substratum  to be
non-relativistic matter ($p \ll \rho$). Then it follows from
(\ref{19}) that
\begin{equation}
\frac{P}{\rho} = \frac{B}{H} + {\cal O}\left(\frac{p}{\rho}\right)\ .
\label{21}
\end{equation}
This may be regarded as the effective equation of state of the
cosmic medium. The quantity $B$, which determines the strength
of the force, is directly related to the effective fluid pressure.
This opens the possibility to establish an explicit relation
between the force function $B$, which quantifies the microscopic
interaction and the cosmological parameters. Namely, comparing the
result (\ref{21}) from kinetic theory for particles in a force
field with Eq.~(\ref{6}) [or (\ref{9})], which is a consequence of
the field equations (\ref{1}), we may simply read off the fraction
$B/H$ which is equivalent to a given value of the deceleration
parameter, namely
\begin{equation}
\frac{B}{H} = - {1 \over 3}\left(1 - 2q \right)
= - \frac{1}{3}\left[\frac{R}{3H^2}- 1\right] \ .
\label{22}
\end{equation}
This relation is the key element of our approach. It relates the
force function $B$ to the Hubble rate and to the deceleration
parameter. Consequently, the effective one-particle force, which
gives rise to a cosmological dynamics, characterized by a Hubble
rate $H$ and a deceleration parameter $q$, is
\begin{eqnarray}
mF^i &=& - \frac{H}{3}\left(1 - 2q \right)
\left[- E p^i + m^2 u^i\right]\nonumber\\
&=& - \frac{H}{3}\left[\frac{R}{3H^2}- 1\right]
\left[- E p^i + m^2 u^i\right]
\ .
\label{23}
\end{eqnarray}
This quantity depends on the microscopic particle momenta but also
on the Hubble rate $H$ and the deceleration parameter $q$. Through
the expression (\ref{23}), the microscopic particle motion,
governed by Eq.~(\ref{16}), is self-consistently coupled to the
cosmological dynamics. The parameters $H$ and $q$ enter the
microscopic dynamics and determine the effective fluid pressure
$\Pi$; in turn, via the field equations (\ref{1}), $\Pi$ is
coupled again to $H$ and $q$. According to
Eqs.~(\ref{6})--(\ref{9}), the force is related to the Ricci
scalar. It is proportional to the deviation from the  flat dust
Universe ($R = 3H^2$). This force describes an interaction of the
individual particle with a space-time curvature, which is
determined by the ensemble of particles itself, i.e. it represents
a curvature self-interaction. All the properties of a force of the
type (\ref{17}) remain valid in this case. In particular, this
self-interaction is compatible with the cosmological principle.
For any given $H$ and $q$ we may construct a force field that
produces the desired dynamics. The described procedure, which
relies on identifying the quantities (\ref{6}) [or (\ref{9})] and
(\ref{21}), couples the gas dynamics self-consistently to the Ricci
scalar, more precisely to the quantity $R - 3H^2$. In a sense,
this may be regarded as a gas-dynamical counterpart to
corresponding couplings of a scalar field to $R$.

Curvature forces are generally not admitted in Einstein's theory
since they represent a non-minimal coupling and violate the
equivalence principle. Here, the mapping of the curvature
interaction on an effective viscous pressure allows a treatment as
an imperfect fluid within the framework of general relativity. We
emphasize that our approach does not introduce new fundamental
particles or fields and preserves Einstein gravity (the left-hand
side of Einstein's equations). It remains open, however, whether
the force (\ref{23}) represents a physical reality or just a
phenomenological fit to some other underlying microphysics. One
might also think of an interpretation according to which averaging
the inhomogeneous matter configuration (see, e.g., \cite{Buchert}
for recent accounts) gives rise to a back-reaction on the
homogeneous background dynamics, such that an epoch of accelerated
expansion is induced by the process of structure formation
\cite{S}. A force of this type, being the result of an
averaging procedure on cosmological scales, would hardly be
detectable in accelerator experiments. 

The force (\ref{23}) may be split into components parallel and perpendicular
to the  comoving velocity:
\begin{equation}
mu_i F^i =
B(E^2 - m^2)\ , \qquad
me_i F^i =  -B
E \sqrt{E^2 - m^2} \ ,
\label{24}
\end{equation}
where
\begin{equation}
e^i \equiv {1\over \sqrt{E^2 - m^2}} \left(p^i - E u^i\right) \
\label{25}
\end{equation}
is the spatial direction of the particle momentum. In the
non-relativistic limit, the spatial projection of the force
becomes
\begin{equation}
e_i F^i \approx - B m v \ .
\label{27}
\end{equation}
For $q<1/2$, the quantity $B = -{1\over 3}H(1-2q)$ plays the role of a
negative friction coefficient. This allows us to interpret the previously
discussed cosmic antifriction \cite{ZSBP} as the result of a non-minimal
coupling of the gas dynamics to the Ricci scalar, equivalent to a specific
curvature self-interaction of the cosmic medium.

\section{Curvature force and accelerated expansion}
\label{Curvature force and accelerated expansion}

So far, we have established a link between the dynamical pressure
$P \approx \Pi$ and the coefficient of antifriction $-B$, and we
have shown that this antifriction can be interpreted as the result
of a non-minimal coupling of matter to curvature. In order to
study the dynamics for a model with given departure from the
Einstein--de Sitter case, we have to integrate the equation
\begin{equation}
\frac{P}{\rho} = - 1 - {2\over3}\frac{\dot{H}}{H^2}
\approx \frac{B}{H}\ ,
\label{28}
\end{equation}
which follows from (\ref{6}) and (\ref{21}). To solve this
equation, an assumption on $B(H,\dot{H})$ is necessary.
Alternatively, one might start from an assumption on the
deceleration parameter $q$  and $B/H$ from Eq.~(\ref{22}).
It is convenient to express the Hubble rate as a function of
redshift $z = \left(a_0/a \right) - 1$. With
\[
\dot{H} = - H^{\prime}H (1 + z) \ ,
\]
where $H^{\prime} \equiv  \mbox{d}H/\mbox{d}z$, the resulting equation is
\begin{equation}
\frac{H ^{\prime }}{B+H}
= \frac{3}{2(1+z)} \ .
\label{29}
\end{equation}

\subsection{Power-law expansion}
\label{Power-law}

The ansatz
\[
B = \sigma \frac{\dot{H}}{H} - \nu H
\]
with constant, non-negative parameters $\sigma$ and $\nu$ leads to
\begin{equation}
H(z) = H_0(1+z)^{1 + q},
\qquad q = {1 - 3(\sigma + \nu)\over 2 + 3 \sigma} \ .
\label{30}
\end{equation}
This is nothing but a power-law expansion, $a \propto t^{n}$, with
$n = 1/(1+q)$. Since $q$ is the (constant) deceleration parameter,
we consistently find that the exponent $n$ is larger than unity
for any $-1 < q <0$, equivalent to the conditions $3(\sigma + \nu)
>1$ and $\nu < 1$. This is the simplest case of a self-consistent
solution of the cosmological dynamics with curvature
self-interaction. Any power $1/(1+q)$ corresponds to a specific
force function $B$ [cf. Eq.~(\ref{22})], with some degeneracy
$\sigma(\nu)$ for a given $q$.

\subsection{The rolling tachyon}
\label{tachyon}

String-theory-inspired tachyon matter was introduced by Sen
\cite{sen}, and its cosmological consequences as an alternative to
a minimally coupled scalar field were explored in
\cite{Gibbons,Pad,Frol,Fein}. A rolling tachyon field $\varphi$
may lead to a power-law behaviour of the scale factor
\cite{Pad,Fein} similar to a scalar field with exponential potential
\cite{Lucchin}. Tachyon matter,
which is characterized by
\begin{equation}
\rho =  \frac{V}{\sqrt{1-\dot{\varphi}^2}} \qquad {\rm and} \qquad
P = - V \sqrt{1-\dot{\varphi}^2} \ , \label{31}
\end{equation}
has recently received some attention as a possible candidate
for dark matter and/or dark energy. Here we demonstrate how the
corresponding dynamics may be related to our present approach.
Assuming $\sigma =0 $, the relevant connection is established by
\[
\nu = 1-\dot{\varphi}^2 \ .
\]
Since $\nu$ is assumed to be constant, $\dot{\varphi}$ has to be
constant as well, which represents a  special case of the tachyon
dynamics, namely $\ddot{\varphi}=0$ and $3HV\dot{\varphi} +
\mbox{d}V /\mbox{d}\varphi = 0$, the quantity $V$ being the
tachyon potential. The corresponding Hubble rate is determined by
\begin{equation}
H^2 = \frac{8\pi G}{3}\rho
= H_0^2(1+z)^{3\dot{\varphi}^2}
 \ .
\label{33}
\end{equation}
It follows that
\begin{equation}
\dot{\rho} = - 3H \rho\dot{\varphi}^2
 \ ,
\label{34}
\end{equation}
which implies
\begin{equation}
\frac{P}{\rho} = - (1-\dot{\varphi}^2) \ . \label{35}
\end{equation}
This is the general equation of state for tachyonic
matter, here obtained for the special case $\varphi \propto
t$ and $V \propto \varphi^{-2}$ \cite{Pad,Fein}, which, according
to (\ref{27}), corresponds to a force field with spatial
projection
\begin{equation}
e_i F^i \approx  (1-\dot{\varphi}^2)H m v \ .
\label{36}
\end{equation}
As was pointed out in \cite{Pad2}, the energy density $\rho$ and
pressure $P$ of the tachyon field may be considered as the sum
of two components according to
\begin{equation}
\rho = \rho_V + \rho_{\rm DM}\ , \qquad P= p_V + p_{\rm DM} \ ,
\end{equation}
where
\begin{equation}
\rho_{\rm DM} =
\frac{V \dot{\varphi}^2}{\sqrt{1-\dot{\varphi}^2}}
\ , \qquad p_{\rm DM} = 0\ ,
\end{equation}
\begin{equation}
\rho_V = V\sqrt{1-\dot{\varphi}^2}
\ , \qquad p_{V} = -\rho_V\ .
\end{equation}
The first component behaves as a pressureless fluid, the second
one has a negative pressure. The power $n$ (recall that $a
\propto t^n$) is then related to the ratio $\rho_V /\rho_{\rm
DM}$ by
\begin{equation}
n = \frac{1}{1+q}
= {2\over 3}\frac{1}{1-\nu} =
{2\over 3}\left(1+\frac{\rho_V}{\rho_{\rm DM}}\right)
\ .
\end{equation}
It follows that this dynamics is realized for a force with
\[
B = - \frac{\rho_V}{\rho}H \ .
\]

\subsection{The $\Lambda$CDM model}
\label{Lambdacdm}

A constant deceleration parameter $q$, equivalent to a constant
ratio $\Pi/\rho$, is not expected to provide a realistic
description of the cosmological dynamics over a large range in
redshift. Successful structure formation requires a period of
decelerated, matter-dominated expansion of substantial length.
Consistent with this requirement, the SNIa data suggest an onset
of accelerated expansion at $z \sim 1$ \cite{Riess01}. (Notice,
however, that there are models that allow structure formation
also during acceleration \cite{amendstruc}. In such a scenario the
accelerated epoch could have started as early as $z\approx 5$
\cite{amendz}). Therefore, a realistic model has to account for
a transition from positive to negative values of the deceleration
parameter. A simple choice which admits this kind of transition is
$|B| \propto H ^{-1}$, equivalent to the ansatz
\begin{equation}
\frac{B}{H} = - \frac{1}{\mu + 1}\frac{H _{0}^{2}}{H ^{2}}\ ,
\label{37}
\end{equation}
where $\mu $ is a constant. Integration of Eq. (\ref{29})  with
the ansatz (\ref{37}) yields
\begin{equation}
H = H_0 \left[\Omega_{\rm CDM}(1+z)^3 + \Omega_{\Lambda}\right]^{1/2}
\label{38}
\end{equation}
with $\Omega_{\rm CDM} = \mu/(\mu + 1)$ and $\Omega_{\Lambda} =1/(\mu
+ 1)$. For $z \gg 1$ we have $H \propto (1+z)^{3/2}$, which is
characteristic of a matter-dominated Universe. For the opposite
case, $z \to -1$, the Hubble rate approaches the constant value $H
\rightarrow H_0 \Omega_{\Lambda}^{1/2}$. The Hubble rate
(\ref{38}) implies a transition from a matter-dominated Universe
at $z\gg 1$ to a de Sitter universe as $z \to -1$. It reproduces
the $\Lambda$CDM model. The observationally favoured value is
$\Omega_{\Lambda} \approx 0.7$. One realizes by direct calculation
that the Hubble rate (\ref{38}) leads to
\begin{equation}
1 - 2q  = 3\frac{\Omega_{\Lambda}}
{\Omega_{\rm CDM} \left(1 + z \right)^{3} + \Omega_{\Lambda}}\ ,
\label{40}
\end{equation}
which is indeed consistent with the general relation (\ref{22}).
For large $z$, we have $q \to 1/2$, while $q \to -1$ for $z \to
-1$. Consequently, the $\Lambda$CDM model is equivalent to a
non-relativistic gas in which a curvature force of the type
(\ref{27}) is self-consistently exerted on the individual
particles. Any ratio $\Omega_{\Lambda}/\Omega_{\rm CDM}$ can be traced
back to a specific curvature self-interaction of the medium. The
explicit expression for the antifriction  coefficient  $|B(z)| =
(1-2q)H/3$ in Eq.~(\ref{27}) is
\begin{equation}
|B(z)|
= H_0 \frac{\Omega_{\Lambda}}
{\left[\Omega_{\rm CDM} \left(1 + z \right)^{3}
+ \Omega_{\Lambda}\right]^{1/2}}
\ .
\label{42}
\end{equation}
For $z \gg 1$ we have $e_{i}F^{i} \rightarrow 0$, which
corresponds to simple, non-interacting dust. For the opposite case
$z \to -1$ the force projection approaches $e_{i}F^{i} \rightarrow
H_0 \Omega_{\Lambda}^{1/2} mv$, with the asymptotic Hubble rate
[cf. Eq. (\ref{38}) for $z \to -1$] as curvature antifriction
constant. In other words, the interaction is gradually switched on
during the cosmic expansion.  

In our approach, the `coincidence
problem', i.e.\ the question, why $\Omega_{\Lambda}$ and
$\Omega_{\rm CDM}$ happen to be of the same order today, is equivalent
to the question: why is the cosmic force parameter $|B(z)|$ of the
order of the Hubble rate just at the present epoch? There are
tentative suggestions that the answer to this question might be
related to the onset of the non-linear stage of the cosmic
structure formation process \cite{S}.

\subsection{The Chaplygin gas}
\label{Chaplygin}

A Chaplygin gas is defined by the equation of state (see
\cite{Kamen,Bilic} and references therein) $P = -A/\rho$,
where $A$ is a positive-definite constant. It can be generalized
by putting an arbitrary power ($\alpha > 0$)
\begin{equation}
P = - \frac{A}{\rho^\alpha} \ . \label{43}
\end{equation}
This equation has the appealing feature of providing a negative
pressure and at the same time a speed of sound that remains real
and positive.  It is reminiscent of certain cases of
string-driven inflation---see Eq.~(2.7) of \cite{Barrow} with
$m<0$. Support for this exotic fluid (with $\alpha =1$) can 
also  be found in higher dimensional theories \cite{Jackiw};
likewise Bento {\it et al.} showed that Eq.~(\ref{43}) can be
derived from a Lagrangian of the Born--Infeld type \cite{Bento}.
Integration of the continuity equation allows us to obtain the
energy density
\begin{equation}
\rho = \left(A + \frac{D}{a^{3(1 +\alpha)}}\right)^{1/(1+\alpha)} \ ,
\label{45}
\end{equation}
where $D$ is a constant. For large values of $a$ the energy
density becomes a cosmological constant. For small values of $a$
it behaves like matter. This property has recently made the
Chaplygin gas an interesting candidate for a one-component model
of the cosmic substratum \cite{Kamen,Bento,Fabris}. However,
new observational constraints seem to restrict the parameter
$\alpha$ to very small values (\cite{Sandvik,Bean}) for which the
Chaplygin gas becomes indistinguishable from the $\Lambda$CDM
model.

Let us now show that in our curvature force approach the
Chaplygin gas can be obtained  with an ansatz $B = -\beta  H^{-2
\alpha - 1}$, where $\beta$ is a non-negative constant. {}From
Eq.~(\ref{21}) and the Friedmann equation, we may immediately read
off that for a choice $\beta = (8\pi G/3)^{(\alpha + 1)}$ we have
\begin{equation}
\frac{P}{\rho} \approx \frac B H = - \frac{A}{\rho^{\alpha + 1}}
\end{equation}
and, consequently
\begin{equation}
B(z) = - \left(8\pi G\over 3\right)^{\alpha + 1}{A \over  H^{2
\alpha + 1}} \ .
\end{equation}
Thus the generalized Chaplygin gas is equivalent to a force
field with spatial projection (according to Eq.~(\ref{27}))
\begin{equation}
e_i F^i \approx \left(8\pi G\over 3\right)^{\alpha + 1} {A m v
\over H^{2 \alpha + 1}} \ .
\label{46}
\end{equation}
We point out that the above relations for the Chaplygin gas
(as well as those for the other examples given in this paper) rely on an
assumption for the dependence $B=B(H)$ of the force function $B$
on the Hubble rate. Via Eq.~(\ref{21}) and the Friedmann equation,
this is equivalent to an effective equation of state $P=P(\rho)$
(see the discussion in the introduction).

\section{Conclusion}
\label{Conclusions}

We have established a scheme that self-consistently relates the
expansion behaviour of the Universe to curvature self-interactions
in the cosmic gas. Assuming that a fluid picture is allowed for
the description of the present Universe, we have constructed
specific internal interactions, which may give rise to the
observed cosmological evolution. The $\Lambda$CDM model may be
regarded as the consequence of a non-relativistic particle motion,
which is non-minimally coupled to the Ricci scalar. This
corresponds to a curvature force, equivalent to a negative
friction, characterized by Eqs.~(\ref{27}) and (\ref{42}).
Alternative dark energy candidates such as a rolling tachyon 
(here with $\phi \propto t$ and $V \propto \phi^{-2}$)  or a Chaplygin
gas have been traced back to curvature interactions in a similar
manner. The corresponding negative friction coefficients are given
by Eqs.~(\ref{36}) and (\ref{46}), respectively. We conclude that
actually any cosmological model with given $H(z)$ and $q(z)$ may
be interpreted on the basis of a gas model with a self-consistent
coupling to the space-time curvature. The presented approach 
rather than introducing new particles or fields introduces a new
effective coupling of gravity to matter. The required
non-minimal interaction can be incorporated into general
relativity. Whether the corresponding force is a physical reality
or the consequence of a back-reaction due to an averaging
procedure, or whether it just provides a phenomenological fit to
some other underlying microphysics, remains open at this stage.
However, contrasting the CMB anisotropies with a perturbation
analysis to be performed elsewhere, is already likely to constrain
the admissible interactions.
\ \\
\ \\
{\bf Acknowledgement}\\
This study was supported by the Deutsche Forschungsgemeinschaft,
the Spanish Ministry of Science and Technology (Grant BFM 2000-0351-C03-01)
and NATO grant PST. CLG.977973.

\end{document}